\documentclass{article}

\usepackage[dblblindworkshop, final]{neurips_2026}

\usepackage[utf8]{inputenc}
\usepackage[T1]{fontenc}
\usepackage{hyperref}
\usepackage{url}
\usepackage{booktabs}
\usepackage{amsfonts}
\usepackage{amsmath}
\usepackage{nicefrac}
\usepackage{microtype}
\usepackage{graphicx}
\usepackage{xcolor}
\usepackage{makecell} % for \Xhline
\usepackage{multirow} % for the multi-backbone results table

\title{Slim Foundation Models for Latency-Critical Recommendation Serving}
\workshoptitle{Long-Context Foundation Models}

\author{%
  Danyal Aftab \\
  Technological University Dublin \\
  Dublin, Ireland \\
  \texttt{Danyal.Aftab@TUDublin.ie}
  \And
  Anas Zafar \\
  National University of Computer and Emerging Science\\
  Karachi, Pakistan \\
  \texttt{anaszafar98@gmail.com}
  \And
  Assadig Abakr \\
  Technological University Dublin \\
  Dublin, Ireland \\
  \texttt{D23125151@mytudublin.ie}
  \And
  Steven Davy \\
  Technological University Dublin \\
  Dublin, Ireland \\
  \texttt{Steven.Davy@TUDublin.ie} \\
}

\begin{document}

\maketitle

\begin{abstract}
Click-through rate (CTR) prediction is a core component of recommendation and
advertising systems. Large language models (LLMs) offer rich semantic
priors that improve prediction quality, particularly in cold-start and
long-tail settings. Injecting a full-scale LLM into a CTR pipeline is,
however, prohibitively expensive at inference time: existing LLM-enhanced CTR
methods avoid this cost by caching LLM outputs offline rather than
serving the LLM online. We propose a resource-efficient pipeline,
\textbf{SlimFM}, that makes online LLM-based semantic modeling
practical: (1) dependency-aware structured pruning compresses LLaMA3-8B by
44\% to a 4.5B-parameter backbone; (2) LoRA-based field-aware fine-tuning
trains the pruned model to produce discriminative field-level semantic
embeddings from lightweight textual prompts; (3) a semantic-guided fusion
module injects these embeddings into the feature interactions of a
conventional CTR model. Across four datasets and six CTR backbones, SlimFM retains over
99\% of uncompressed-model accuracy while reducing inference latency by up to
$3.4\times$ relative to existing LLM-enhanced baselines.
\end{abstract}

\section{Introduction}
\label{sec:intro}

CTR prediction is a critical component in recommender systems and digital
advertising \citep{guo2017deepfm,zhou2018deep,gligorijevic2019deeply}.
Large language models can enrich this pipeline with semantic and world
knowledge that ID-based models lack, particularly for cold-start items and
long-tail users \citep{xi2024towards,li2023ctrl,qiu2024ease}. The primary obstacle is computational cost: modern CTR backbones comprise on the order of 1--10M parameters, while
contemporary LLMs such as LLaMA3-8B comprise 8B or more, making inference
with a complete LLM infeasible under real-time serving constraints. Prior LLM-enhanced CTR methods therefore pre-compute and cache LLM outputs
offline, serving only the CTR model online.
Representative methods such as KAR \citep{xi2024towards}, LLM-CF
\citep{sun2024large}, CTRL \citep{li2023ctrl}, and EASE \citep{qiu2024ease}
incur, on average, over 290 times higher computational cost than standard CTR
baselines \citep{cui2025fieldmatterslightweightllmenhanced}. This approach
transfers the computational burden from online serving to offline preprocessing
rather than eliminating it; moreover, cached outputs must be regenerated
whenever the item or feature catalog changes. This work investigates whether structured pruning of the LLM, combined with
parameter-efficient fine-tuning, can compress the backbone sufficiently to
enable online serving within the same request path as the CTR model, without
significant degradation in semantic quality.

\textbf{Contributions.}
\begin{itemize}
  \item A dependency-aware structured pruning procedure that compresses
  LLaMA3-8B to a 4.5B-parameter backbone (44\% smaller) while preserving the
  representational capacity needed for downstream semantic-embedding
  extraction.
  \item Field-aware LoRA-based fine-tuning and a semantic-guided
  fusion mechanism together enable a pruned LLM to be served online
  rather than cached offline inside a latency-critical CTR pipeline.
  \item Across four datasets $\times$ six CTR backbones, 44\% parameter reduction
  retains $>$99\% of baseline accuracy at up to $3.4\times$ lower inference
  latency than existing LLM-enhanced CTR baselines.
\end{itemize}

\section{Method}
\label{sec:method}

\begin{figure}[t]
  \centering
  \includegraphics[width=\linewidth]{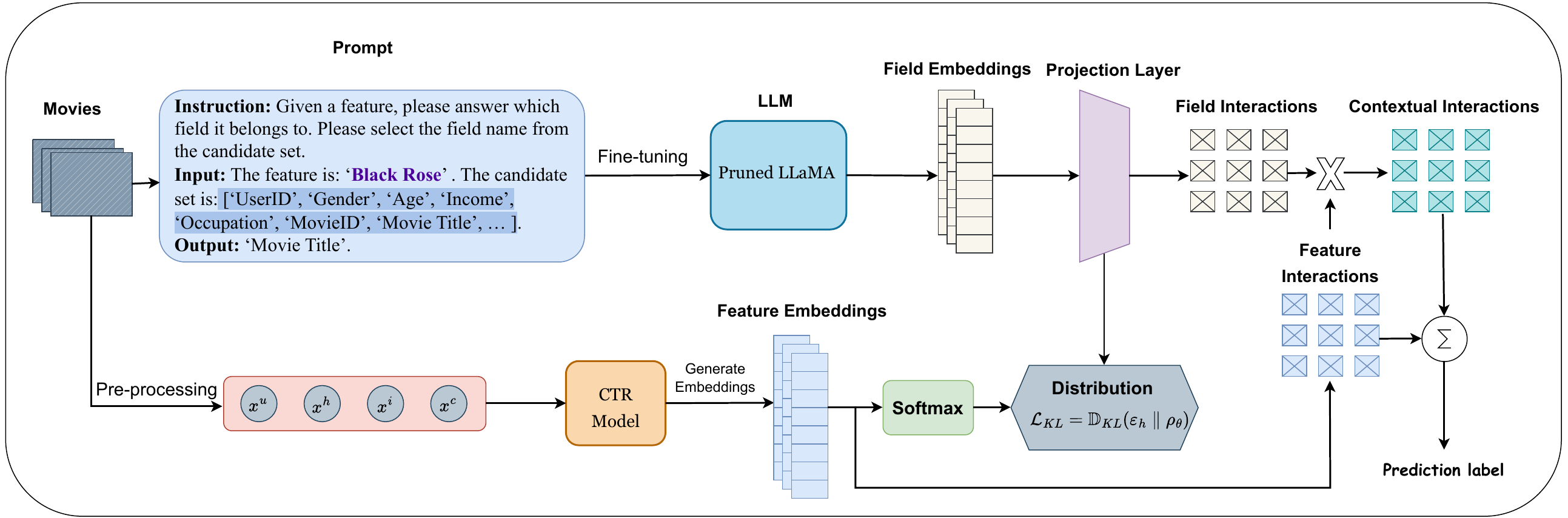}
  \caption{The SlimFM framework. The upper branch fine-tunes a pruned LLaMA
  with feature-field prompts to extract field embeddings, which are projected
  into pairwise field interactions. The lower branch generates feature
  embeddings from the CTR model and computes feature interactions. The two
  streams are fused via element-wise multiplication and aggregated for
  prediction, with a KL-divergence loss aligning the embedding distributions
  during training.}
  \label{fig:framework}
  
\end{figure}

Figure~\ref{fig:framework} summarizes the three stages of SlimFM.

\paragraph{Structured Model Pruning.}
Following DepGraph \citep{fang2023depgraph} and LLM-Pruner
\citep{ma2023llmpruner}, we build a dependency graph over the attention and
feed-forward parameters of LLaMA3-8B \citep{dubey2024llama}, grouping
structurally coupled parameters into prunable units $\mathcal{G} =
\{P_k\}_{k=1}^N$. The importance of each group is estimated via a Taylor expansion of the
next-token prediction loss:
\begin{equation}
  I_{\mathcal{G}} = \sum_{k=1}^{N} \left| \frac{\partial \mathcal{L}}{\partial P_k}^{\!\top} P_k - \tfrac{1}{2} P_k^{\top} H P_k \right|,
  \label{eq:importance}
\end{equation}
where $H$ is the Hessian. Groups with the lowest $I_{\mathcal{G}}$ are removed
until a target sparsity is reached, yielding a 4.5B-parameter backbone.

\paragraph{Field-Aware Semantic Learning.}
We adapt the pruned LLM to the CTR domain via a field-identification task,
using structured prompts that list the $M$ candidate field names
(e.g., classifying \texttt{`Black Rose'} as field \texttt{MovieID}). For a feature value $v$ from field $f$,
the LLM encodes the prompt into a query representation $h$ and each
candidate field name into embeddings $\{e_{f_j}\}_{j=1}^{M}$. We train the model using an InfoNCE-style contrastive loss:
\begin{equation}
  \mathcal{L}_{\text{field}} = -\log \frac{\exp(h \cdot e_{f^{+}}/\tau)}{\sum_{j=1}^{M} \exp(h \cdot e_{f_j}/\tau)},
  \label{eq:contrastive}
\end{equation}
where $f^{+}$ is the correct field label and $\tau$ is a temperature
hyperparameter. Only the LoRA adapters \citep{hu2022lora} on the
attention projection matrices are updated; the pruned backbone itself remains frozen.

\paragraph{Semantic-Guided Feature Fusion.}
After fine-tuning, we extract one embedding $e_f$ per field by encoding the
field name and projecting the final hidden state of the model. Pairwise cosine
similarities $S^{\text{llm}}_{ij}$ between field embeddings are passed
through a learnable transformation $g_\theta$ to obtain task-adapted weights
$\tilde{S}^{\text{llm}}_{ij}$, which reweight the pairwise feature interactions
of the CTR model:
\begin{equation}
  C^{\text{llm}}_{ij} = \tilde{S}^{\text{llm}}_{ij} \cdot I^{\text{ctr}}_{ij},
  \label{eq:fusion}
\end{equation}
where $I^{\text{ctr}}_{ij} = e^{\text{ctr}}_i \cdot e^{\text{ctr}}_j$ denotes
the pairwise CTR feature interaction. These reweighted interactions are
aggregated into a contextual prediction stream
$y_{\text{context}} = \sum_{i<j} C^{\text{llm}}_{ij}$, then combined with
the interaction stream of the CTR model. During training, a
KL-divergence additionally aligns the pairwise similarity distributions
induced by the CTR and LLM embedding spaces. The overall loss combines
prediction and alignment, $\mathcal{L} = \mathcal{L}_{\text{BCE}} +
\lambda_{\text{align}} \mathcal{L}_{\text{align}}$, where $\mathcal{L}_{\text{BCE}}$
is the binary cross-entropy loss for click prediction. At inference time, the pruned backbone is deployed directly, enabling
online LLM-based semantic modeling without pre-caching.

\section{Experiments}
\label{sec:experiments}

\begin{table}[!htb]
  \centering
  \small
  \begin{tabular}{lcccc}
    \toprule
     & Digital Music & MovieLens-1M & Video Games & Gift Cards \\
    \midrule
    \#Field    & 13 & 8 & 13 & 13 \\
    \#Feature  & 1.70M & 9.00M & 60.1M & 1.98M \\
    \#User     & 101K & 6.04K & 2.77M & 133K \\
    \#Item     & 70.5K & 3.71K & 137K & 1.14M \\
    \#Interact.& 130K & 100M & 4.62M & 152K \\
    Density    & $1.83\times10^{-5}$ & $4.47\times10^{-2}$ & $1.22\times10^{-5}$ & $1.01\times10^{-3}$ \\
    \bottomrule
  \end{tabular}
  \caption{Dataset statistics.}
  \label{tab:datasets}
\end{table}

\begin{table}[!htb]
  \centering
  {\footnotesize \setlength{\tabcolsep}{0.99mm}
  \resizebox{1\linewidth}{!}{%
  \begin{tabular}{c|ccc|ccc|ccc|ccc|ccc|ccc}
    \Xhline{1.2pt}
    & \multicolumn{3}{c|}{FM} & \multicolumn{3}{c|}{DeepFM} & \multicolumn{3}{c|}{FwFM} & \multicolumn{3}{c|}{FmFM} & \multicolumn{3}{c|}{FinalMLP} & \multicolumn{3}{c}{Wukong} \\
    Method & Par. & MAC & Lat. & Par. & MAC & Lat. & Par. & MAC & Lat. & Par. & MAC & Lat. & Par. & MAC & Lat. & Par. & MAC & Lat. \\
    & (M) & (G) & (ms) & (M) & (G) & (ms) & (M) & (G) & (ms) & (M) & (G) & (ms) & (M) & (G) & (ms) & (M) & (G) & (ms) \\
    \hline
    KAR & 8032 & 996 & 54.62 & 8032 & 996 & 54.71 & 8032 & 996 & 54.78 & 8032 & 996 & 54.81 & 8033 & 996 & 54.95 & 8034 & 996 & 55.08 \\
    LLM-CF & 7075 & 829 & 46.56 & 7075 & 829 & 46.46 & 7075 & 829 & 47.02 & 7075 & 829 & 47.02 & 7076 & 829 & 47.43 & 7077 & 829 & 47.53 \\
    \textbf{SlimFM (Ours)} & \textbf{4504} & \textbf{275} & \textbf{15.99} & \textbf{4504} & \textbf{275} & \textbf{15.98} & \textbf{4504} & \textbf{275} & \textbf{16.89} & \textbf{4504} & \textbf{275} & \textbf{17.58} & \textbf{4504} & \textbf{275} & \textbf{16.41} & \textbf{4505} & \textbf{275} & \textbf{16.42} \\
    \Xhline{1.2pt}
  \end{tabular}
  }}
  \caption{Computational efficiency comparison of LLM-enhanced CTR methods}
  \label{tab:efficiency}
  % \vspace{-5em}
\end{table}

\begin{table}[!h]
\centering
{\footnotesize \setlength{\tabcolsep}{0.9mm}
\resizebox{1\linewidth}{!}{%
\begin{tabular}{cccccccccccccccc}
\Xhline{1.2pt}
\multicolumn{2}{c|}{} & \multicolumn{3}{c|}{Gift Cards} & \multicolumn{3}{c|}{Video Games} & \multicolumn{3}{c|}{Digital Music} & \multicolumn{3}{c|}{MovieLens-1M} &
Mean  \% \\
\cline{3-14}
\multicolumn{2}{c|}{\multirow{-2}{*}{Method}} & LL $\downarrow$ & AUC $\uparrow$ & \multicolumn{1}{c|}{RelImp} & LL $\downarrow$ & AUC $\uparrow$ & \multicolumn{1}{c|}{RelImp} & LL $\downarrow$ & AUC $\uparrow$ & \multicolumn{1}{c|}{RelImp} & LL $\downarrow$ & AUC $\uparrow$ & \multicolumn{1}{c|}{RelImp} & AUC $\uparrow$  \\ \hline
\multicolumn{2}{c|}{CELA} & 0.4658 & 0.6746 & \multicolumn{1}{c|}{1.02\%} & 0.5798 & 0.6775 & \multicolumn{1}{c|}{0.44\%} & 0.3364 & 0.7638 & \multicolumn{1}{c|}{0.37\%} & 0.4031 & 0.8373 & \multicolumn{1}{c|}{0.24\%} & 0.7383  \\
\multicolumn{2}{c|}{ClickPrompt} & 0.4656 & 0.6754 & \multicolumn{1}{c|}{1.48\%} & 0.5747 & 0.6799 & \multicolumn{1}{c|}{1.79\%} & 0.3356 & 0.7652 & \multicolumn{1}{c|}{0.90\%} & 0.4025 & 0.8379 & \multicolumn{1}{c|}{0.42\%} & 0.7396 \\ \hline
\multicolumn{1}{c|}{} & \multicolumn{1}{c|}{Base} & 0.4606 & 0.6621 & \multicolumn{1}{c|}{} & 0.5805 & 0.6673 & \multicolumn{1}{c|}{} & 0.3359 & 0.7612 & \multicolumn{1}{c|}{} & 0.4081 & 0.8354 & \multicolumn{1}{c|}{} & 0.7315  \\
\multicolumn{1}{c|}{} & \multicolumn{1}{c|}{KAR} & 0.4551 & 0.6643 & \multicolumn{1}{c|}{1.34\%} & 0.5803 & 0.6705 & \multicolumn{1}{c|}{1.93\%} & 0.3410 & 0.7607 & \multicolumn{1}{c|}{-0.21\%} & 0.4073 & 0.8358 & \multicolumn{1}{c|}{0.13\%} & 0.7328   \\
\multicolumn{1}{c|}{} & \multicolumn{1}{c|}{LLM-CF} & 0.4580 & 0.6666 & \multicolumn{1}{c|}{2.77\%} & 0.5805 & 0.6682 & \multicolumn{1}{c|}{0.58\%} & \textbf{0.3351} & \textbf{0.7633} & \multicolumn{1}{c|}{\textbf{0.78\%}} & 0.4128 & 0.8337 & \multicolumn{1}{c|}{-0.50\%} & 0.7330   \\
\multicolumn{1}{c|}{} & \multicolumn{1}{c|}{CTRL} & 0.4595 & 0.6636 & \multicolumn{1}{c|}{0.93\%} & \textbf{0.5802} & \textbf{0.6708} & \multicolumn{1}{c|}{\textbf{2.13\%}} & 0.3360 & 0.7611 & \multicolumn{1}{c|}{-0.04\%} & 0.4128 & 0.8336 & \multicolumn{1}{c|}{-0.53\%} & 0.7323   \\
\multicolumn{1}{c|}{} & \multicolumn{1}{c|}{EASE} & \textbf{0.4534} & \textbf{0.6667} & \multicolumn{1}{c|}{\textbf{2.84\%}} & 0.5804 & 0.6701 & \multicolumn{1}{c|}{1.69\%} & 0.3359 & 0.7625 & \multicolumn{1}{c|}{0.48\%} & 0.4101 & 0.8357 & \multicolumn{1}{c|}{0.10\%} & \textbf{0.7338}   \\
\multicolumn{1}{c|}{\multirow{-7}{*}{FM}} & \multicolumn{1}{c|}{SlimFM (Ours)} & 0.4790 & 0.6644 & \multicolumn{1}{c|}{1.42\%} & 0.5804 & 0.6691 & \multicolumn{1}{c|}{1.08\%} & 0.3360 & 0.7620 & \multicolumn{1}{c|}{0.31\%} & \textbf{0.4049} & \textbf{0.8360} & \multicolumn{1}{c|}{\textbf{0.15\%}} & 0.7329    \\ \hline
\multicolumn{1}{c|}{} & \multicolumn{1}{c|}{Base} & 0.4751 & 0.6713 & \multicolumn{1}{c|}{} & 0.5762 & 0.6692 & \multicolumn{1}{c|}{} & 0.3478 & 0.7582 & \multicolumn{1}{c|}{} & 0.4190 & 0.8342 & \multicolumn{1}{c|}{} & 0.7332  \\
\multicolumn{1}{c|}{} & \multicolumn{1}{c|}{KAR} & \textbf{0.4753} & \textbf{0.6778} & \multicolumn{1}{c|}{\textbf{3.80\%}} & 0.5871 & 0.6720 & \multicolumn{1}{c|}{1.65\%} & 0.3439 & 0.7587 & \multicolumn{1}{c|}{0.18\%} & 0.4207 & 0.8335 & \multicolumn{1}{c|}{-0.20\%} & 0.7355 \\
\multicolumn{1}{c|}{} & \multicolumn{1}{c|}{LLM-CF} & 0.4723 & 0.6736 & \multicolumn{1}{c|}{1.36\%} & 0.5769 & 0.6730 & \multicolumn{1}{c|}{2.28\%} & 0.3441 & 0.7588 & \multicolumn{1}{c|}{0.21\%} & 0.4140 & 0.8353 & \multicolumn{1}{c|}{0.34\%} & 0.7352   \\
\multicolumn{1}{c|}{} & \multicolumn{1}{c|}{CTRL} & 0.4746 & 0.6744 & \multicolumn{1}{c|}{1.83\%} & \textbf{0.5757} & \textbf{0.6758} & \multicolumn{1}{c|}{\textbf{3.92\%}} & 0.3432 & 0.7576 & \multicolumn{1}{c|}{-0.26\%} & 0.4142 & 0.8352 & \multicolumn{1}{c|}{0.31\%} & 0.7358   \\
\multicolumn{1}{c|}{} & \multicolumn{1}{c|}{EASE} & 0.4748 & 0.6739 & \multicolumn{1}{c|}{1.52\%} & 0.5807 & 0.6738 & \multicolumn{1}{c|}{2.74\%} & 0.3360 & 0.7604 & \multicolumn{1}{c|}{0.83\%} & 0.4180 & 0.8349 & \multicolumn{1}{c|}{0.22\%} & 0.7358   \\
\multicolumn{1}{c|}{\multirow{-7}{*}{DeepFM}} & \multicolumn{1}{c|}{SlimFM (Ours)} & 0.4565 & 0.6724 & \multicolumn{1}{c|}{0.64\%} & 0.5856 & 0.6716 & \multicolumn{1}{c|}{1.42\%} & \textbf{0.3326} & \textbf{0.7686} & \multicolumn{1}{c|}{\textbf{4.03\%}} & \textbf{0.4217} & \textbf{0.8356} & \multicolumn{1}{c|}{\textbf{0.42\%}} & \textbf{0.7371}  \\ \hline
\multicolumn{1}{c|}{} & \multicolumn{1}{c|}{Base} & 0.4664 & 0.6691 & \multicolumn{1}{c|}{} & 0.5762 & 0.6683 & \multicolumn{1}{c|}{} & 0.3420 & 0.7553 & \multicolumn{1}{c|}{} & 0.4017 & 0.8365 & \multicolumn{1}{c|}{} & 0.7323  \\
\multicolumn{1}{c|}{} & \multicolumn{1}{c|}{KAR} & 0.4661 & 0.6694 & \multicolumn{1}{c|}{0.19\%} & 0.5791 & 0.6741 & \multicolumn{1}{c|}{3.42\%} & 0.3432 & 0.7575 & \multicolumn{1}{c|}{0.88\%} & 0.4020 & 0.8360 & \multicolumn{1}{c|}{-0.15\%} & 0.7343  \\
\multicolumn{1}{c|}{} & \multicolumn{1}{c|}{LLM-CF} & 0.4672 & 0.6698 & \multicolumn{1}{c|}{0.47\%} & 0.5806 & 0.6702 & \multicolumn{1}{c|}{1.13\%} & 0.3397 & 0.7567 & \multicolumn{1}{c|}{0.57\%} & 0.4044 & 0.8371 & \multicolumn{1}{c|}{0.18\%} & 0.7335   \\
\multicolumn{1}{c|}{} & \multicolumn{1}{c|}{CTRL} & 0.4679 & 0.6699 & \multicolumn{1}{c|}{0.50\%} & 0.5776 & 0.6755 & \multicolumn{1}{c|}{4.28\%} & 0.3414 & 0.7561 & \multicolumn{1}{c|}{0.33\%} & \textbf{0.4004} & \textbf{0.8372} & \multicolumn{1}{c|}{\textbf{0.21\%}} & 0.7347  \\
\multicolumn{1}{c|}{} & \multicolumn{1}{c|}{EASE} & 0.4661 & 0.6703 & \multicolumn{1}{c|}{0.74\%} & 0.5795 & 0.6765 & \multicolumn{1}{c|}{4.84\%} & 0.3415 & 0.7571 & \multicolumn{1}{c|}{0.71\%} & 0.4030 & 0.8369 & \multicolumn{1}{c|}{0.12\%} & 0.7352  \\
\multicolumn{1}{c|}{\multirow{-7}{*}{FwFM}} & \multicolumn{1}{c|}{SlimFM (Ours)} & \textbf{0.4652} & \textbf{0.6714} & \multicolumn{1}{c|}{\textbf{1.36\%}} & \textbf{0.5745}  & \textbf{0.6774} & \multicolumn{1}{c|}{\textbf{5.41\%}} & \textbf{0.3342} & \textbf{0.7621} & \multicolumn{1}{c|}{\textbf{2.66\%}} & 0.4017 & 0.8366 & \multicolumn{1}{c|}{0.03\%} & \textbf{0.7369}  \\ \hline
\multicolumn{1}{c|}{} & \multicolumn{1}{c|}{Base} & 0.4579 & 0.6747 & \multicolumn{1}{c|}{} & 0.5824 & 0.6706 & \multicolumn{1}{c|}{} & 0.3315 & 0.7628 & \multicolumn{1}{c|}{} & 0.4077 & 0.8325 & \multicolumn{1}{c|}{} & 0.7352 \\
\multicolumn{1}{c|}{} & \multicolumn{1}{c|}{KAR} & \textbf{0.4595}  & \textbf{0.6754} & \multicolumn{1}{c|}{\textbf{0.42\%}} & 0.5767 & 0.6711 & \multicolumn{1}{c|}{0.29\%} & 0.3347 & 0.7640 & \multicolumn{1}{c|}{0.43\%} & 0.4062 & 0.8332 & \multicolumn{1}{c|}{0.21\%} & 0.7359   \\
\multicolumn{1}{c|}{} & \multicolumn{1}{c|}{LLM-CF} & 0.4652 & 0.6740 & \multicolumn{1}{c|}{-0.43\%} & 0.5877 & 0.6709 & \multicolumn{1}{c|}{0.20\%} & 0.3317 & 0.7627 & \multicolumn{1}{c|}{-0.04\%} & 0.4066 & 0.8329 & \multicolumn{1}{c|}{0.11\%} & 0.7351  \\
\multicolumn{1}{c|}{} & \multicolumn{1}{c|}{CTRL} & 0.4593 & 0.6751 & \multicolumn{1}{c|}{0.21\%} & 0.5826 & 0.6714 & \multicolumn{1}{c|}{0.47\%} & \textbf{0.3310} & \textbf{0.7650} & \multicolumn{1}{c|}{\textbf{0.82\%}} & 0.4059 & 0.8332 & \multicolumn{1}{c|}{0.21\%} & 0.7362  \\
\multicolumn{1}{c|}{} & \multicolumn{1}{c|}{EASE} & 0.4590 & 0.6753 & \multicolumn{1}{c|}{0.33\%} & 0.5750 & 0.6729 & \multicolumn{1}{c|}{1.35\%} & 0.3311 & 0.7637 & \multicolumn{1}{c|}{0.33\%} & \textbf{0.4056} & \textbf{0.8334} & \multicolumn{1}{c|}{\textbf{0.27\%}} & \textbf{0.7363}  \\
\multicolumn{1}{c|}{\multirow{-7}{*}{FmFM}} & \multicolumn{1}{c|}{SlimFM (Ours)} & 0.4596 & 0.6710 & \multicolumn{1}{c|}{-2.12\%} & \textbf{0.5731} & \textbf{0.6759} & \multicolumn{1}{c|}{\textbf{3.11\%}} & 0.3327 & 0.7645 & \multicolumn{1}{c|}{0.65\%} & 0.4057 & 0.8333 & \multicolumn{1}{c|}{0.24\%} & 0.7362 \\ \hline
\multicolumn{1}{c|}{} & \multicolumn{1}{c|}{Base} & 0.4774 & 0.6720 & \multicolumn{1}{c|}{} & 0.5830 & 0.6761 & \multicolumn{1}{c|}{} & 0.3423 & 0.7557 & \multicolumn{1}{c|}{} & 0.4086 & 0.8301 & \multicolumn{1}{c|}{} & 0.7335 \\
\multicolumn{1}{c|}{} & \multicolumn{1}{c|}{KAR} & 0.4881 & 0.6726 & \multicolumn{1}{c|}{0.35\%} & \textbf{0.5820} & \textbf{0.6817} & \multicolumn{1}{c|}{\textbf{3.19\%}} & 0.3429 & 0.7546 & \multicolumn{1}{c|}{-0.41\%} & 0.4094 & 0.8290 & \multicolumn{1}{c|}{-0.33\%} & 0.7345   \\
\multicolumn{1}{c|}{} & \multicolumn{1}{c|}{LLM-CF} & 0.4781 & 0.6731 & \multicolumn{1}{c|}{0.62\%} & 0.5845 & 0.6785 & \multicolumn{1}{c|}{1.36\%} & 0.3369 & 0.7560 & \multicolumn{1}{c|}{0.13\%} & 0.4041 & 0.8299 & \multicolumn{1}{c|}{-0.06\%} & 0.7344  \\
\multicolumn{1}{c|}{} & \multicolumn{1}{c|}{CTRL} & 0.4719 & 0.6781 & \multicolumn{1}{c|}{3.55\%} & 0.5835 & 0.6790 & \multicolumn{1}{c|}{1.65\%} & \textbf{0.3399}  & \textbf{0.7561} & \multicolumn{1}{c|}{\textbf{0.15\%}} & \textbf{0.4033} & \textbf{0.8328} & \multicolumn{1}{c|}{\textbf{0.82\%}} & \textbf{0.7365}   \\
\multicolumn{1}{c|}{} & \multicolumn{1}{c|}{EASE} & 0.4773 & 0.6736 & \multicolumn{1}{c|}{0.93\%} & 0.5776 & 0.6796 & \multicolumn{1}{c|}{1.97\%} & 0.3369 & 0.7560 & \multicolumn{1}{c|}{0.12\%} & 0.4049 & 0.8309 & \multicolumn{1}{c|}{0.24\%} & 0.7350  \\
\multicolumn{1}{c|}{\multirow{-7}{*}{FinalMLP}} & \multicolumn{1}{c|}{SlimFM (Ours)} & \textbf{0.4666} & \textbf{0.6792} & \multicolumn{1}{c|}{\textbf{4.19\%}} & 0.5945 & 0.6733 & \multicolumn{1}{c|}{-1.59\%} & 0.3443 & 0.7553 & \multicolumn{1}{c|}{-0.16\%} & 0.4170 & 0.8251 & \multicolumn{1}{c|}{-1.51\%} & 0.7332 \\ \hline
\multicolumn{1}{c|}{} & \multicolumn{1}{c|}{Base} & 0.4755 & 0.6728 & \multicolumn{1}{c|}{} & 0.5819 & 0.6767 & \multicolumn{1}{c|}{} & 0.3402 & 0.7514 & \multicolumn{1}{c|}{} & 0.4088 & 0.8323 & \multicolumn{1}{c|}{} & 0.7333 \\
\multicolumn{1}{c|}{} & \multicolumn{1}{c|}{KAR} & 0.4794 & 0.6721 & \multicolumn{1}{c|}{-0.43\%} & 0.5927 & 0.6758 & \multicolumn{1}{c|}{-0.53\%} & 0.3429 & 0.7506 & \multicolumn{1}{c|}{-0.31\%} & 0.4099 & 0.8316 & \multicolumn{1}{c|}{-0.21\%} & 0.7325   \\
\multicolumn{1}{c|}{} & \multicolumn{1}{c|}{LLM-CF} & 0.4756 & 0.6752 & \multicolumn{1}{c|}{1.36\%} & 0.5805 & 0.6788 & \multicolumn{1}{c|}{1.15\%} & 0.3379 & 0.7560 & \multicolumn{1}{c|}{1.83\%} & 0.4084 & 0.8293 & \multicolumn{1}{c|}{-0.90\%} & 0.7348  \\
\multicolumn{1}{c|}{} & \multicolumn{1}{c|}{CTRL} & 0.4753 & 0.6769 & \multicolumn{1}{c|}{2.37\%} & 0.5792 & 0.6800 & \multicolumn{1}{c|}{1.84\%} &  \textbf{0.3399} & \textbf{0.7561} & \multicolumn{1}{c|}{\textbf{1.85\%}} & 0.4029 & 0.8367 & \multicolumn{1}{c|}{1.31\%} & \textbf{0.7374}   \\
\multicolumn{1}{c|}{} & \multicolumn{1}{c|}{EASE} & 0.4748 & 0.6758 & \multicolumn{1}{c|}{1.71\%} & \textbf{0.5783}  & \textbf{0.6801} & \multicolumn{1}{c|}{\textbf{1.91\%}} & 0.3381 & 0.7556 & \multicolumn{1}{c|}{1.67\%} & 0.4031 & 0.8334 & \multicolumn{1}{c|}{0.33\%} & 0.7362  \\
\multicolumn{1}{c|}{\multirow{-7}{*}{Wukong}} & \multicolumn{1}{c|}{SlimFM (Ours)} & \textbf{0.4539} & \textbf{0.6808} & \multicolumn{1}{c|}{\textbf{4.63\%}} & 0.5830 & 0.6744 & \multicolumn{1}{c|}{-1.30\%} & 0.3463 & 0.7535 & \multicolumn{1}{c|}{0.84\%} & \textbf{0.4028}  & \textbf{0.8371} & \multicolumn{1}{c|}{\textbf{1.44\%}} & 0.7365  \\ \hline
\multicolumn{2}{c|}{\textbf{Average RelImp}} & \multicolumn{3}{c|}{\textbf{1.69\%}} & \multicolumn{3}{c|}{\textbf{1.36\%}} & \multicolumn{3}{c|}{\textbf{1.39\%}} & \multicolumn{3}{c|}{\textbf{0.13\%}} &   \\ \hline
\Xhline{1.2pt}
\end{tabular}
}}
\caption{Performance comparisons of SlimFM with existing LLM-enhanced CTR methods across six backbone models and four datasets. SlimFM uses a pruned LLaMA3 (4.5B parameters). LL denotes Logloss and RelImp denotes Relative Improvement. \textbf{Bold} values indicate the best result within each backbone group.}
\label{tab:performance1}
\end{table}

We evaluate SlimFM on four public CTR datasets: Amazon Gift Cards, Amazon Video
Games, Amazon Digital Music, and MovieLens-1M, as shown in Table~\ref{tab:datasets}. We use six CTR backbones: FM
\citep{rendle2010factorization}, DeepFM \citep{guo2017deepfm}, FwFM
\citep{pan2018field}, FmFM \citep{sun2021fm2}, FinalMLP
\citep{mao2023finalmlp}, and Wukong \citep{zhang2024wukong}. We compare
against LLM-enhanced baselines KAR \citep{xi2024towards}, LLM-CF
\citep{sun2024large}, CTRL \citep{li2023ctrl}, and EASE \citep{qiu2024ease},
all using the same LLaMA3-8B backbone for a fair comparison, plus two PLM-enhanced baselines, CELA \citep{wang2024cela} and
ClickPrompt \citep{lin2024clickprompt}.

\textbf{Efficiency.} Table~\ref{tab:efficiency} reports parameters,
multiply-accumulate operations (MACs), and single-sample inference latency
(averaged over 50 runs with 10 warmup runs) across all six CTR backbones.
The pruned backbone of SlimFM comprises \textbf{4504M} parameters,
substantially fewer than KAR (8032M) and LLM-CF (7075M), representing a $1.8\times$
reduction. This compression translates directly into lower inference
latency: SlimFM consistently achieves \textbf{15.98--17.58\,ms} per
prediction, versus \textbf{54.62--55.08\,ms} for KAR and
\textbf{46.46--47.53\,ms} for LLM-CF, a $3.1$--$3.4\times$ speedup over KAR
and $2.7$--$2.9\times$ over LLM-CF that holds across all six backbones.
The near-invariance of per-method latency to CTR backbone choice confirms that
the LLM forward pass is the inference bottleneck; structured pruning of the LLM
is therefore the critical factor for reducing end-to-end inference latency.

\textbf{Accuracy.} Table~\ref{tab:performance1} compares SlimFM against
traditional CTR models, PLM-enhanced methods, and
LLM-enhanced methods across all six backbones and
four datasets. Averaged over the six backbones, SlimFM improves AUC relative
to the base CTR model of each backbone by \textbf{1.69\%} on
Gift Cards, \textbf{1.36\%} on Video Games, \textbf{1.39\%} on Digital Music,
and \textbf{0.13\%} on MovieLens-1M. Notable combinations include Wukong on Gift Cards (AUC \textbf{0.6808}, +4.63\%)
and FinalMLP on Gift Cards (\textbf{0.6792}, +4.19\%), both the highest AUC among
LLM-enhanced methods in their respective backbone group; FwFM achieves the
largest gain on Video Games (AUC \textbf{0.6774}, +5.41\%); and DeepFM the
strongest improvement on Digital Music (AUC \textbf{0.7686}, +4.03\%). Overall, SlimFM
achieves the highest AUC among KAR, LLM-CF, CTRL, and EASE in 10 of the 24
backbone/dataset combinations; its RelaImpr remains positive across all datasets
despite a 44\% smaller LLM backbone than the uncompressed baselines.

\begin{table}[h]
  \centering
  \small
  \resizebox{\linewidth}{!}{%
  \begin{tabular}{lcccccccccc}
    \toprule
    Sparsity & \multicolumn{2}{c}{Gift Cards} & \multicolumn{2}{c}{Video Games} & \multicolumn{2}{c}{Digital Music} & \multicolumn{2}{c}{MovieLens-1M} & Mean & Recovery \\
     & LogLoss & AUC & LogLoss & AUC & LogLoss & AUC & LogLoss & AUC & AUC & Rate \\
    \midrule
    LLaMA3-8B    & 0.4500 & 0.6673 & 0.5791 & 0.6718 & 0.3346 & 0.7629 & 0.4040 & 0.8364 & 0.7346 & -- \\
    LLaMA3-6B    & 0.4666 & 0.6656 & 0.5798 & 0.6703 & 0.3351 & 0.7625 & 0.4044 & 0.8362 & 0.7337 & 99.88\% \\
    LLaMA3-4.5B  & 0.4790 & 0.6644 & 0.5804 & 0.6691 & 0.3357 & 0.7620 & 0.4049 & 0.8360 & 0.7329 & 99.77\% \\
    LLaMA3-2B    & 0.4997 & 0.6623 & 0.5816 & 0.6672 & 0.3359 & 0.7619 & 0.4053 & 0.8357 & 0.7318 & 99.62\% \\
    \bottomrule
  \end{tabular}
  }
  \caption{Sparsity-performance trade-off analysis for SlimFM with LLaMA3 backbones.}
  \label{tab:sparsity}
\end{table}
\textbf{Sparsity-performance trade-off.} Table~\ref{tab:sparsity} reports
accuracy across pruning levels for the FM backbone, ranging from
8B model (mean AUC \textbf{0.7346}) to 2B parameters.
Pruning to 6B retains \textbf{99.88\%} of the mean AUC, and pruning to 4.5B retains \textbf{99.77\%} while
achieving roughly $1.8\times$ compression; only at the more aggressive 2B
setting does the recovery rate drop further, to \textbf{99.62\%}, with the
decline most visible on Gift Cards and Video Games. The compression trade-off is smooth and monotone, with no sharp degradation
point, indicating that the 4.5B configuration offers a favorable balance
between model compression and predictive accuracy.

\section{Conclusion}
\label{sec:conclusion}

We show that dependency-aware structured pruning of LLaMA3-8B, combined with
field-aware fine-tuning and semantic-guided fusion, enables online LLM-based
semantic modeling within a latency-critical CTR pipeline while retaining over
99\% of full-model accuracy. These results demonstrate that structured pruning
is not merely an optimization but a necessary mechanism for practical
LLM-enhanced CTR deployment. Our evaluation is restricted to four CTR datasets and a single LLM family (LLaMA3-8B), so generalization to other backbones, domains, and LLM architectures remains to be verified.

\begin{ack}
This publication has emanated from research conducted with the financial support of Taighde Eireann -- Research Ireland under Grant number 21/FFP-A/9174.
\end{ack}

\bibliographystyle{plain}
\bibliography{references}

\newpage

\end{document}